# Effects of carbon concentration and filament number on advanced internal-Mg-infiltration-processed MgB$_2$ Strands

G Z Li[1], M D Sumption[1], J B Zwayer[1], M A Susner[1], M A Rindfleisch[2], C J Thong[2], M J Tomsic[2] and E W Collings[1]

[1] Center for Superconducting and Magnetic Materials, Department of Materials Science and Engineering, the Ohio State University, Columbus, OH 43210, U.S.A.
[2] Hyper Tech Research Incorporated, 539 Industrial Mile Road, Columbus, OH 43228, U.S.A.

**Abstract**

An advanced internal Mg infiltration method (AIMI) in this paper has been shown to be effective in producing superconducting wires containing dense MgB$_2$ layers with high critical current densities. In this study, the in-field critical current densities of a series of AIMI-fabricated MgB$_2$ strands were investigated in terms of C doping levels, heat treatment (HT) time and filament numbers. The highest layer $J_c$ for our monofilamentary AIMI strands is $1.5 \times 10^5$ A/cm$^2$ at 10 T, 4.2 K, when the C concentration was 3 mol% and the strand was heat-treated at 675 °C for 4 hours. Transport critical currents were also measured at 4.2 K on short samples and one-meter segments of eighteen-filament C-doped AIMI strands. The layer $J_c$s reached $4.3 \times 10^5$ A/cm$^2$ at 5 T and $7.1 \times 10^4$ A/cm$^2$ at 10 T, twice as high as those of the best PIT strands. The analysis of these results indicates that the AIMI strands, possessing both high layer $J_c$s and engineering $J_e$s after further optimization, have strong potential for commercial applications.

## 1. Introduction

Extensive efforts have been made to improve the transport properties of superconductors based on $MgB_2$ since the discovery of its superconducting properties in 2001 [1]. Typical $MgB_2$ powder-in-tube (PIT) wires, consisting of an $MgB_2$ core surrounded by a chemical barrier and a hard outer sheath, can be made by either the *ex situ* or the *in situ* routes [2]. Through the *ex situ* technique, it is possible to develop homogeneous $MgB_2$ wires with high powder packing densities over long lengths [3]. Since the powder fill consists of pre-reacted $MgB_2$, the wires can be used either as-formed or after a sintering heat treatment (HT) at temperatures of around 800-900 °C [4, 5]. The *in situ* wires need to be heat-treated in order to react the mixed Mg and B powders. Typical HT temperatures are 650-800 °C [6, 7] although lower temperatures, even below the melting point of Mg, have been used [8]. High porosity and weak connectivity are critical issues associated with conventional *in situ* PIT $MgB_2$ wires [9], even though they have some of the highest $J_c$ values present in the literatures [10, 11].

An interesting variant of the *in situ* PIT route is the "internal Mg diffusion" (IMD) process or the "reactive liquid-Mg infiltration" (Mg-RLI) process initiated by Giunchi *et al.* [12, 13]. The conductor is formed by Mg from the central rod diffusing into the surrounding B precursors in certain conditions and then reacting into a $MgB_2$ hollow cylinder. Whereas the PIT process produces a porous $MgB_2$ core, the IMD process provides a dense $MgB_2$ layer with excellent longitudinal and transverse connectivities [14-16]. To calculate the critical current densities of these IMD wires, it is important to take care with the definition of the areas to which the critical current, $I_c$, is normalized. Three areas are commonly used depending on different purposes [17]: (a) the layer critical current density, layer $J_c$, is defined by using the cross-sectional area of $MgB_2$ in the composite. All the other components, including the central hole in IMD wires, are ignored; (b) the non-barrier critical current density, non-barrier $J_c$, takes into consideration the cross-sectional area of everything within the chemical barrier. It is best used for direct quality comparison between different types of $MgB_2$ wires; (c) the engineering critical current density, $J_e$, adopts the cross-sectional area of the entire strand. This is an important parameter for engineering designs.

Numerous attempts have been made to develop high performing IMD wires, including doping [18], adding extra Mg into B layers to assist the $MgB_2$ layer growth [19], optimizing HT conditions [20], and adjusting filament numbers and wire geometries [21, 22]. Adding different kinds of chemicals into wires, such as SiC [21] and liquid aromatic hydrocarbon [23], were proved to achieve layer $J_c$s and engineering $J_e$s, which are much higher than undoped IMD wires. For example, Kumakura *et al.* [21] fabricated a series of 10 mol% SiC doped $MgB_2$ wires using the IMD process. A high layer $J_c$ of $1.1 \times 10^5$ A/cm$^2$ was attained at 4.2 K and 10 T for one of their monofilamentary wires HTed at 600 °C. By using C doped nano-sized amorphous B powders, our group also made IMD wires with a good layer $J_c$ of $1.0 \times 10^5$ A/cm$^2$ obtained at 4.2 K and 10 T [16]. Also, Togano *et al.* [20] prepared a series of multifilamentary IMD wires. They considered the nineteen-filamentary IMD wires as optimal candidates for high performing strands, because the fine filaments in the nineteen-filamentary IMD wires enabled a thinner $MgB_2$ layer and hence were more suitable to gain full $MgB_2$ phase transformation. A good layer $J_c$ of $9.9 \times 10^4$ A/cm$^2$ was obtained for their nineteen-filamentary IMD wire at 4.2 K and 10 T. However, given the low "$MgB_2$ fill factor", (i.e. the effective $MgB_2$ cross sectional area fraction in the whole strand), this wire only achieved the engineering $J_e$ of less than $3 \times 10^3$ A/cm$^2$ at 4.2 K and 10 T, even though the B in it has been nearly fully transformed into MgB$_2$.

Nevertheless, because the maximum MgB$_2$ layer thickness was always limited to 20-30 $\mu$m [20], the $J_e$s of all of these IMD wires were actually lower than state of the art PIT wires. To overcome the limitations of MgB$_2$ layer thickness in previous IMD wires, our group studied the MgB$_2$ layer growth mechanism associated with the Mg diffusion route. In our recent report [24], our efforts in choosing optimized B powder types, various wires diameters, HT conditions and wire constructions finally leaded to high performing wires with maximum layer $J_c$ of $1.1 \times 10^5$ A/cm$^2$ and maximum engineering $J_e$ of $1.7 \times 10^4$ A/cm$^2$ at 4.2 K and 10 T. The $J_e$s for those samples were now much higher than those of best of class PIT strands. To indicate the substantial advances in the capabilities of these "second generation" MgB$_2$ strands, and to point to the large quantitative difference in the practical properties between these conductors and previous wires, we describe these "$J_e$-optimized" strands as "advanced internal Mg infiltration" (AIMI) wires.

Although it was widely reported that dopants, especially C and carbides, were beneficial in enhancing the layer $J_c$ and engineering $J_e$, the effect of the C doping level on these properties has not been systematically investigated in either early IMD-synthesized wires or our new AIMI wires. So it is necessary to figure out the optimal C concentration for the AIMI wires. Moreover, considering the excellent layer $J_c$ and engineering $J_e$ properties of monocore AIMI strands, it is also essential to develop multifilamentary AIMI wires for application. In addition, most reports about the diffusion or infiltration-processed wires to date have focused on short wires (e.g. << 1 m total length); only a few measurements have been made on long strands [22] although a one-meter-long undoped Mg-RLI monofilament wire fabricated by Giunchi *et al.* had a engineering $J_e$ of over 1.0

× $10^4$ A/cm$^2$ at 4.2 K and 5 T [25]. Though the AIMI wires have higher layer $J_c$s than their PIT counterparts, many other characteristics such as engineering $J_e$, MgB$_2$ fill factor and thermal stability need to be optimized in long wires before this manufacturing process is applied commercially.

For this study a group of 3 mol% and 4 mol% C doped monofilamentary "advanced internal Mg infiltration-processed" (AIMI) strands have been fabricated and compared with the previously reported 2 mol% C monocore strand. The critical transport currents were measured and the results were reported in terms of layer $J_c$s and non-barrier $J_c$s after the appropriate cross sectional areas were measured by scanning electron microscopy (SEM). The effects of C doping level on the layer $J_c$s and microstructures of the monofilamentary wires have been studied. Multifilamentary strands of eighteen sub-elements were also prepared using the same C-doping concentrations. Their respective values of layer $J_c$, non-barrier $J_c$ and engineering $J_e$ were then compared with a set of conventional multifilamentary PIT strands at 4.2 K. In particular, the 2 mol% C doped eighteen-filamentary AIMI wires were made into one-meter-long strands for the purpose of large scale characterization of the superconducting properties of the AIMI strands.

## 2. Experimental

*2.1 Sample preparation*

In this paper, three groups of MgB$_2$ strands, including (i) monofilamentary AIMI strands, (ii) multifilamentary AIMI strands and (iii) typical multifilamentary PIT strands, were manufactured by Hyper Tech Research Inc. (HTR). All strands used the C-doped B powder produced by the plasma assisted reaction between $BCl_3$, $H_2$ and suitable amount of $CH_4$ (Specialty Materials Inc., SMI) [26]. This powder was mostly amorphous, 10-100 nm in size, with C doping levels ranging from 2 mol% to 4 mol%.

The wire fabrication procedure is described as follow: (i) The strand geometry of the monofilamentary AIMI strands were based on our previous studies of 2 mol% C doped "2G-IMD" strands [24] and 2-4 mol% C doped PIT strands [27]. The starting billet was a Mg rod positioned along the axis of a B-filled double tube of Nb and Monel; the B is doped with either 3 mol% or 4 mol% C. Then the billet was drawn to 0.55 mm outer diameter (OD). (ii) To fabricate multifilamentary AIMI strands, the initial billet, of the same geometry as that of monofilamentary AIMI strands, was drawn to an intermediate-sized monofilament. Then the eighteen of these monofilaments and one central Cu-Ni alloy filament were restacked into a Monel tube, and further drawn to an OD of 0.83 mm. (iii) To compare with multifilamentary AIMI strands in terms of critical current densities, a group of thirty-six-filamentary PIT strands with variable C doping levels were fabricated by HTR, using the "continuous tube filling-and–forming (CTFF)" approach [28]. As before the starting monofilamentary PIT strands were drawn to smaller diameter. Then thirty six as-drawn monofilaments together with a central Cu filament were restacked into a Monel tube and further drawn to 0.92 mm OD prior to HT.

After wire drawing, the three sets of strands were placed in a tube furnace respectively, ramped to soak temperature in about 80 min, kept at certain temperature for one, two or four hours and furnace cooled to room temperature. Since in this study we mainly considered the effect of HT time on $MgB_2$ wires, the HT temperature of 675 °C was chosen for all AIMI strands and most PIT strands. In fact, 675 °C is often regarded as the optimal temperature for CTFF-type PIT wires and little variation in $J_c$ was found with HT temperature at 650-700 °C for suitable times. Furthermore, all HT time was below 4 hours because longer sintering time would cause $MgB_2$ grain coarsening which were harmful for $MgB_2$ conductors. The strand specifications and HT conditions of the above mentioned three groups of strands are listed in table 1.

*2.2 Characterization*

Transport voltage-current measurements were performed on all samples at 4.2 K in pool boiling liquid He in transverse magnetic fields, *B*, of up to 13.5 T. Two types of samples were studied in this work: (1) "Short" straight samples 50 mm long, with a gauge length of about 5 mm. (2) "ITER-barrel" samples made with one-meter-long segments helically wound onto 32-mm-diameter Ti-Al-V alloy holders [28]. The gauge length was 500 mm. In both cases, the transport critical current was determined under an electric field criterion of 1 $\mu$V/cm.

The microstructures of the $MgB_2$ wires, including the areas of the reacted $MgB_2$ layers, were characterized by scanning electron microscopy (SEM) and compositions quantified using energy dispersive X-ray spectroscopy (EDS). SEM observation was carried out using FEI (Philips) Sirion field-emission source SEM and Quanta 200 SEM equipped with the EDAX EDS system. Microstructures were also characterized using Olympus PME-3 optical microscope (OM) to obtain better contrast between $MgB_2$ and B layers.

**3. Results**

*3.1 Monofilamentary AIMI strands*

Figure 1 shows the layer $J_c$s of the 3 mol% and 4 mol% C doped monofilamentary $MgB_2$ strands at 4.2 K in magnetic field of 5-13.5 T. A previously reported 2 mol% C doped "2G-IMD" strand HTed at 675 °C for 1 hour [24] is also included in this plot for comparison. As the C doping level increases from 2 mol%, the layer $J_c$s of the AIMI strands are firstly enhanced at 3 mol% C and then suppressed when the C concentration is 4 mol%. The 3 mol% C doped sample A3 achieves the highest layer $J_c$ of $1.5 \times 10^5$ A/cm$^2$ at 10 T. Also, it is noted that the layer $J_c$s of all samples, except A5, increase as the strands are heat treated longer. Take the 3 mol% C doped strands for example, after 1 hour HT, the layer $J_c$ at 10 T is $1.3 \times 10^5$ A/cm$^2$ for A1. It increases to $1.4 \times 10^5$ A/cm$^2$ after 2 hours HT and finally attains $1.5 \times 10^5$ A/cm$^2$ after 4 hours HT.

Although moderately increasing C concentration from 2 mol% to 3 mol% is effective in improving the layer $J_c$, the transverse cross sectional view of the strands shows that the $MgB_2$ layer growth is suppressed for 3 mol% C doped strands and thus the $MgB_2$ layer thickness becomes relatively narrower than the 2 mol% C strands. Figure 2 shows the OM and back scattered SEM images of strand A3. They are taken from the same transverse cross section of the wire. Because of better contrast, the OM picture is helpful to discern the $MgB_2$ and B-rich layers. As indicated in figure 2(a), the orange (or purple) annulus is the $MgB_2$ layer and the outside dark area is the B-rich region. The $MgB_2$ layer looks dense but its layer thickness is limited – only 8-25 $\mu$m, which is less than a quarter of the maximum layer thickness of the previously reported 2 mol% C sample [24] under the same HT condition. The suppression of the $MgB_2$ layer growth caused by the carbide doping has also been reported by other groups [21], but the underlying mechanism is still unclear. Consequently not only the area of $MgB_2$ and central Mg but also the area occupied by the non-superconducting B-rich region is taken into account as the total area to calculate the non-barrier $J_c$s of the wires. Therefore the non-barrier $J_c$s of these samples are reduced compared with the fully reacted AIMI strands.

Figure 3 shows the field dependence of the non-barrier $J_c$s for all of the monofilamentary strands. Sample A3 obtains the highest non-barrier $J_c$ of $2.4 \times 10^4$ A/cm$^2$ at 10 T. This value is lower than that of the fully reacted 2 mol% C doped AIMI strand, but is still comparable to the non-barrier $J_c$s of most PIT wires [16]. Given the high layer $J_c$s of the 3 mol% C doped AIMI strands, there might be a great improvement in non-barrier $J_c$s once the full $MgB_2$ reaction could be realized.

*3.2 Multifilamentary AIMI and PIT strands*

Figure 4(a), the transverse cross-sectional SEM image of strand B1, represents the strand geometry of all multifilamentary AIMI samples. In the transverse direction, all subfilaments are about 100 $\mu$m in size and uniformly deformed. Mg and dispersed powders are found existing in the "prior-Mg" holes in the centers of filaments. Between the Nb barriers and the holes are the reacted $MgB_2$ layers with thickness of 0-30 $\mu$m. After increasing the contrast and brightness of figure 4(a) by 50% and thus the contrast difference within the reaction layers, it is found that only in some filaments has the B powder become fully transformed to $MgB_2$. This is exemplified by the OM picture (Figure 4b), in which some filaments have grey B-rich areas outside the yellow $MgB_2$ circulars, indicating partially transformation. The structure of these multifilamentary strands is also examined longitudinally. Tubular $MgB_2$ layers are seen attached to the Nb

barriers in each filament. These MgB$_2$ filaments are uniform in diameter along the longitudinal direction. However, Nb barrier breakages can be seen at some locations. It is also noted that for every multifilamentary AIMI strand some residual Mg remains in the region of the break. EDS, employed to analyze the composition of the remaining Mg region for strand B3 and B5, detects the presence of Mg and Cu in the atomic ratio of about 4:5, implying that Cu has leaked into the filaments.

The areas of the MgB$_2$ layers are measured based on the transverse SEM images. Since both B and Mg are light atoms, the weak contrast in SEM figures between MgB$_2$ and other borides or B makes measurement difficult even when the contrast and brightness of the pictures are adjusted. So OM images, with better phase contrast, are used to assist distinguishing the MgB$_2$ region. Table 2 lists the MgB$_2$ areas as well as the MgB$_2$ fill factors of all multifilamentary AIMI samples obtained using this method. It is important to note that the MgB$_2$ fill factors of the present AIMI strands are only 9.2 to 14.6 %, significantly lower than those of PIT strands.

The field dependencies of the layer $J_c$s at 4.2 K for all the multifilamentary AIMI strands are shown in figure 5. Those of the two best performing PIT strands, P3 and P4, are also included for comparison. The best performing strand in magnetic field below 10 T is the 2 mol% C doped strand B1. At 5 T its layer $J_c$ reaches $4.3 \times 10^5$ A/cm$^2$; at 10 T it is $7.1 \times 10^4$ A/cm$^2$ which is twice as high as that achieved by the best PIT strand (i.e. the 3 mol% C-doped strand P3). The 4 mol% C doped strand B5 exhibits the highest layer $J_c$s at high fields above 10 T, although its low-field $J_c$s are lower than those of other strands. Its layer $J_c$ attains $2.8 \times 10^4$ A/cm$^2$ at 13 T. Compared with B1 and B5, the 3 mol% C doped strand B3 shows good layer $J_c$s in all applied fields. But unlike its monofilamentary counterpart A1, a slightly heavier doping in multifilamentary AIMI strands does not result in an appreciable layer $J_c$ enhancement as compared with 2 mol% C doped AIMI wires. Secondly, it is noticeable that all the 2-hours-HT strands (i.e. B2, B4 and B6) show deteriorated layer $J_c$s as compared to their 1-hour-HT counterparts B1, B3 and B5 respectively. For example, after two hours HT, the layer $J_c$ of what then becomes strand B2 decreases to $3.3 \times 10^5$ A/cm$^2$ at 5 T and $4.7 \times 10^4$ A/cm$^2$ at 10 T. However, the layer $J_c$s of B2 are still 30% higher than the best PIT results. Also this 1 m long ITER-barrel mounted AIMI strand shows comparable layer $J_c$s with other AIMI short samples, indicating the strand is uniform in the wire direction over long lengths, and it is capable of carrying a transport current of over 1000 A at 1 T, exhibiting excellent thermal stability. Figure 6 shows the non-barrier $J_c$ vs. $\mu_0 H$ curves of the multifilamentary AIMI wires.

For the magnet builder the transport property of importance is the engineering $J_e$. Figure 7 presents $J_e$ as a function of applied field strength. Since $J_e$ is the critical current, $I_c$, divided by the transverse cross-sectional area of the whole wire, it also satisfies the relationship $J_e$ = layer $J_c \times$ MgB$_2$ fill factor. The $J_e$ of the best multifilamentary AIMI strand, B1, is $5.7 \times 10^4$ A/cm$^2$ at 5 T and $9.4 \times 10^3$ A/cm$^2$ at 10 T. Driven by their greater fill factors the best PIT results are here shown to be comparable to those of the best AIMI strands.

## 4. Discussion

The C or carbide doping concentration has been reported to have a great impact on the critical current densities of the conventional MgB$_2$ PIT wires [29, 30]. This effect still works for the monofilamentary AIMI strands. The monocore AIMI strands achieve maximum layer $J_c$s when the C doping concentration is in the vicinity of 3 mol%. The layer $J_c$ is suppressed when the C doping level deviates from this optimal value. This is in agreement with our group's previous results that the PIT strands behave best when the 3 mol% C is mixed into the wires [16, 27]. Proper carbon doping is proved to enhance the upper critical field $B_{c2}$, decrease the anisotropy ratio $\gamma$ and provide extra flux pinning centers [27, 31, 32], all of which are helpful to obtain higher layer $J_c$s. Nevertheless, if too much C is doped, it might deteriorate the connectivity and thus reduce the transport layer $J_c$s [33]. As a consequence, relatively low layer $J_c$s are obtained for 4 mol% C samples when the C doping level is above 3 mol%. Compared with monofilamentary AIMI strands, the multifilamentary AIMI strands possess finer subfilaments and thus are more easily to suffer from degradations during the process of wire fabrication. The Nb barrier breakage observed in B1 and Cu leakage detected in samples B3 and B5 would cause a certain level of deterioration in layer $J_c$s, so the layer $J_c$s of all multifilamentary AIMI strands with any C doping level are lower than those of their monofilamentary counterparts. For the same reason, these extrinsic defects such as Cu contamination have so negative effects on multifilamentary wires that no obvious layer $J_c$ improvement is observed even when the C doping level is optimal (i.e. 3 mol%).

The effect of HT time on the layer $J_c$ is quite opposite for monofilamentary and multifilamentary AIMI strands. A longer HT time, such as 4 hours, is desired for monofilamentary AIMI strands, while a shorter period of HT time is more preferred in multifilamentary strands. The average B precursor layer thickness in multifilamentary strands is less than 20 $\mu$m, so the whole layer quickly transforms into MgB$_2$ within 1 hour during the HT. Extending HT time is not helpful for MgB$_2$ layer growth but could only cause grain coarsening, oxidation or Cu contamination, which are harmful for the strands [34]. Contrarily, the average layer thickness of B precursor in monofilamentary AIMI strands is thicker than 50 $\mu$m. So Mg is not able to diffuse deeply into the B layer within a short time. According to the MgB$_2$ layer growth mechanism in AIMI wires discussed in the early paper [24], a partially reacted MgB$_2$ layer is formed at the very beginning of HT, so moderate $J_c$s are obtained for A1. As HT continues, the MgB$_2$ layer becomes thicker, better transformed and well-connected. Consequently, A2 and A3 achieve improved $J_c$s and $J_e$s.

The above difference also implies that multifilamentary AIMI MgB$_2$ wires are more advantageous to obtain high layer $J_c$s and engineering $J_e$s than monocore AIMI wires. They could be fully reacted within a short time, because the B precursor layers become narrower and thus the Mg diffusion distance is shortened. As the HT is completed faster, the grain coarsening can be prevented, and other detrimental effects like oxidation could be limited to a low level. Also, since the maximum MgB$_2$ layer thickness is always limited to a very small value (for example, 25 $\mu$m for 3 mol% C doped monofilamentary AIMI strands), it is desirable to reduce the thickness of the B precursor layer below the 25 $\mu$m, which can be realized in multifilamentary strands.

The 10 T data of Figures 5 and 7 are combined in Figure 8. Given that layer $J_c \times FF = J_e$ (where $FF$ represents the MgB$_2$ fill factor) plots of layer $J_c$ vs. $FF$ are rectangular hyperbolae corresponding to various selected values of $J_e$. For any given value of $J_e$ there exists a reciprocal relationship between layer $J_c$ and $FF$ which leads in this case to a clustering of the AIMI data around high layer $J_c$ and low $FF$ and conversely for the PIT data. This occurs for the following reason: the MgB$_2$ core of the PIT strands, which occupies some 25% of the strand cross section, being formed by the *in situ* reaction of Mg and B powders is porous. As a result of this, and the fact that the reacted MgB$_2$ exists in the form of imperfectly connected stringers [35], the longitudinal critical current density (the layer $J_c$ in the case of PIT) is relatively low as seen in the figure. In the AIMI strand the cylindrical $MgB_2$ layer is dense, well connected, and free of pores [22]. This leads to a high layer $J_c$. On the other hand the relative volume of the reacted layer (the AIMI strand's *FF*) tends to be small. It is possible, therefore for the AIMI data and the PIT data to cluster around a common $J_e$ curve, but different segments of it. This tendency is seen for AIMI strands B5, B6, and B2 and PIT strands P1, P2, and P5. What Figure 8 tells us, however, is that having optimized the starting powders (Mg plus fine SMI B doped with 2-3 mol% C) and after improving strand processing to eliminate breakages and barrier leakages, a significant increase in $J_e$ to much more than 10 kA/cm$^2$ at 4.2 K, 10 T, will accompany an increase in the *FF* of the AIMI strand.

## 5. Summary

The in-field critical current densities of a series of advanced internal Mg infiltration-processed $MgB_2$ strands have been investigated in terms of C doping level, HT time and filament number. The best monofilamentary AIMI strand A3, with 3 mol% C and under 4 hours HT at 675 °C, achieves high layer $J_c$ of $1.5 \times 10^5$ A/cm$^2$ and non-barrier $J_c$ of $2.4 \times 10^4$ A/cm$^2$ at 10 T and 4.2 K. In multifilamentary AIMI strands, we have not retained the full layer $J_c$ increase due to some defect structure. Even so, their layer $J_c$s, non-barrier $J_c$s and engineering $J_e$s are still better than, or at least comparable to, those of the best PIT strands. The one-meter-long version of the multifilamentary strands show that this kind of advanced $MgB_2$ strands is uniform and thermally stable, capable of being utilized in industry.

**Acknowledgements**

This work is funded by the Department of Energy, High Energy Physics division under Grant No. DE-FG02-95ER40900, a DOE SBIR, and a program from the Ohio Department of Development. The authors thank Dr. Yi Ding from Southeast University for his assistance in sample characterization and helpful discussions.

**List of Tables**

Table 1. Strand Specifications and Heat Treatment Conditions.

Table 2. MgB$_2$ Areas and MgB$_2$ Fill Factors of Multifilamentary AIMI Strands.

Table 1. Strand Specifications and Heat Treatment Conditions [a]

| Sample name | Trace No. [b] | Sample type [c] | Filament count | Central filament | C mol% [d] | MgB$_2$ FF (%) [e] | Filament to strand fraction (%) [f] | MgB$_2$ (%) in filament [g] | dia., mm | H.T., °C /min |
|---|---|---|---|---|---|---|---|---|---|---|
| *Monofilamentary AIMI Samples* | | | | | | | | | | |
| A1 | 2893I | short | 1 | - | 3 | 4.2 | 33.5 | 12.5 | 0.55 | 675/60 |
| A2 | 2893II | short | 1 | - | 3 | 5.1 | 32.2 | 15.8 | 0.55 | 675/120 |
| A3 | 2893IV | short | 1 | - | 3 | 5.2 | 32.5 | 16.0 | 0.55 | 675/240 |
| A4 | 2909I | short | 1 | - | 4 | 3.7 | 31.8 | 11.0 | 0.55 | 675/60 |
| A5 | 2909II | short | 1 | - | 4 | 5.0 | 32.6 | 15.3 | 0.55 | 675/120 |
| A6 | 2909IV | short | 1 | - | 4 | 5.0 | 32.5 | 15.4 | 0.55 | 675/240 |
| *Multifilamentary AIMI Samples* | | | | | | | | | | |
| B1 | 2885I | barrel | 18 | Cu-Ni | 2 | 13.2 | 33.8 | 39.1 | 0.83 | 675/60 |
| B2 | 2885II | barrel | 18 | Cu-Ni | 2 | 14.4 | 33.2 | 43.4 | 0.83 | 675/120 |
| B3 | 2962I | short | 18 | Cu-Ni | 3 | 13.1 | 33.0 | 39.7 | 0.83 | 675/60 |
| B4 | 2962II | short | 18 | Cu-Ni | 3 | 13.0 | 33.6 | 38.7 | 0.83 | 675/120 |
| B5 | 2967I | short | 18 | Cu-Ni | 4 | 9.8 | 33.4 | 29.3 | 0.83 | 675/60 |
| B6 | 2967II | short | 18 | Cu-Ni | 4 | 12.0 | 34.4 | 34.9 | 0.83 | 675/120 |
| *Multifilamentary PIT Samples* | | | | | | | | | | |
| P1 | 2555 | barrel | 36 | Cu | 2 | 25.0 | 25.0 | 100 | 0.92 | 700/60 |
| P2 | 2510 | barrel | 36 | Cu | 2.5 | 22.6 | 22.6 | 100 | 0.92 | 675/60 |
| P3 | 2580 | barrel | 36 | Cu | 3 | 17.8 | 17.8 | 100 | 0.91 | 650/120 |
| P4 | 2538 | barrel | 36 | Cu | 3.5 | 25.8 | 25.8 | 100 | 0.91 | 675/60 |
| P5 | 2535 | barrel | 36 | Cu | 4 | 26.9 | 26.9 | 100 | 0.91 | 675/60 |

[a] All the strands, fabricated by HTR, incorporated a Nb chemical barrier and Monel outer sheath.

[b] Sample number for internal purpose.

[c] As explained in Section 2.2, "barrel" means a one-meter-long segment helically wound onto 32-mm-diameter Ti-Al-V holder; "short" means a 50 mm long straight wire.

[d] C doping level, which is based on C analysis by the LECO Corporation and normalized to the molar weight of MgB$_2$. No assumptions are made here concerning the expected uptake of C into the B sublattice; see, [27].

[e] Namely the MgB$_2$ fill factor, which is the transverse cross-sectional area fraction of MgB$_2$ in the whole strand, based on scanning electron microscopy (SEM) and optical microscopy (OM) images. Hence, MgB$_2$ FF = MgB$_2$ layer area / overall wire area. The "layer $J_c$" is estimated according to "MgB$_2$ FF".

[f] The "Filament to strand fraction" in the non-barrier area fraction in the whole strand, i.e. "Filament to strand fraction" = the cross-sectional area of everything within the Nb chemical barrier area (not including Nb) / overall wire area.

[g] The "MgB$_2$ (%) in filament" = MgB$_2$ layer area / the cross-sectional area of everything within the Nb chemical barrier area (not including Nb).

Table 2. MgB$_2$ Areas and MgB$_2$ Fill Factors of Multifilamentary AIMI Strands.

| Sample name | MgB$_2$ area, $\mu m^2$ | error, $\mu m^2$ + / - | MgB$_2$ fill factor, % | error, % + / - |
|---|---|---|---|---|
| B1 | 71600 | 2700 / 2200 | 13.2 | 0.5 / 0.4 |
| B2 | 77800 | 3900 / 1200 | 14.4 | 0.7 / 0.2 |
| B3 | 70800 | 700 / 1500 | 13.1 | 0.1 / 0.3 |
| B4 | 70500 | 1300 / 3700 | 13.0 | 0.2 / 0.7 |
| B5 | 52900 | 3100 / 300 | 9.8 | 0.6 / 0.1 |
| B6 | 65400 | 2500 / 2500 | 12.0 | 0.5 / 0.5 |

**List of Figures**

Figure 1 Field dependence of the layer $J_c$s of the 3 mol% and 4 mol% C doped monofilamentary AIMI strands at 4.2 K. Also included is a previously reported 2 mol% C doped "2G-IMD" strand HTed at 675 °C for 1 hour [24].

Figure 2 (a) Optical microscopic and (b) back scattered SEM images taken from the same transverse cross section in the 3 mol% C doped AIMI strand A3.

Figure 3 Field dependence of the non-barrier $J_c$s of the 3 mol% and 4 mol% C doped monofilamentary AIMI strands at 4.2 K. A fully transformed 2 mol% C doped "2G-IMD" strand is included [24].

Figure 4 Transverse cross-sectional (a) SEM and (b) OM images of the 2 mol% C doped multifilamentary AIMI strand B1.

Figure 5 Field dependence of the layer $J_c$s for multifilamentary AIMI strands at 4.2 K, data from P3 and P4 are included for comparison. Here B1 and B2 are one-meter-long "ITER-barrel" type strands.

Figure 6 Field dependence of the non-barrier $J_c$s for multifilamentary AIMI strands at 4.2 K, data from P3 and P4 are included for comparison. Here B1 and B2 are one-meter-long "ITER-barrel" type strands.

Figure 7 Field dependence of the engineering $J_e$s for multifilamentary AIMI strands at 4.2 K, data from P3 and P4 are included for comparison. Here B1 and B2 are one-meter-long "ITER-barrel" type strands.

Figure 8 Comparison and relationships between $J_e$, layer $J_c$ and $MgB_2$ fill factor (*FF*) of multifilamentary AIMI strands and PIT strands with different C concentrations at 4.2 K, 10 T. Error bars are added for AIMI strands to describe the accuracy of the data.

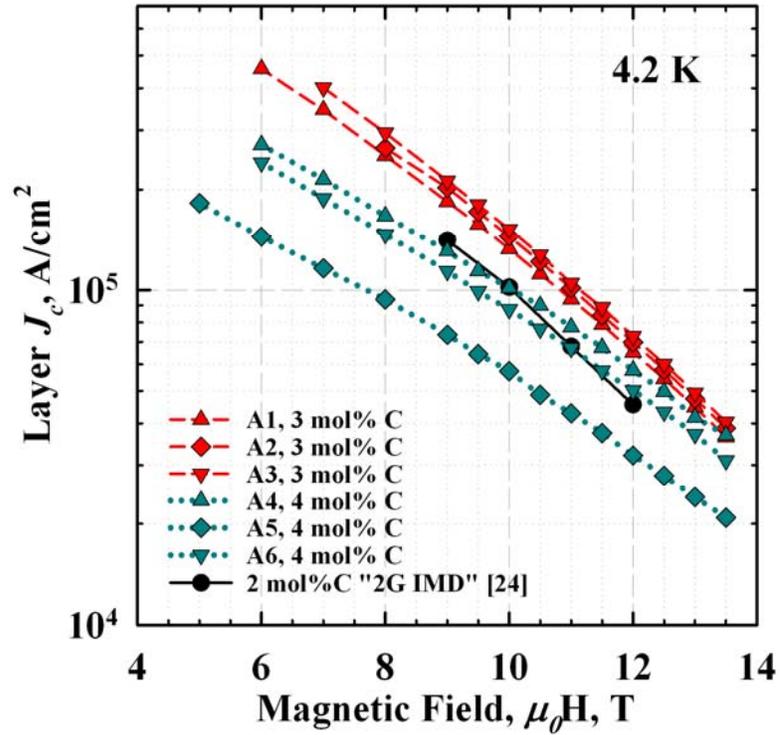

Figure. 1 Field dependence of the layer $J_c$s of the 3 mol% and 4 mol% C doped monofilamentary AIMI strands at 4.2 K. Also included is a previously reported 2 mol% C doped "2G-IMD" strand HTed at 675 °C for 1 hour [24].

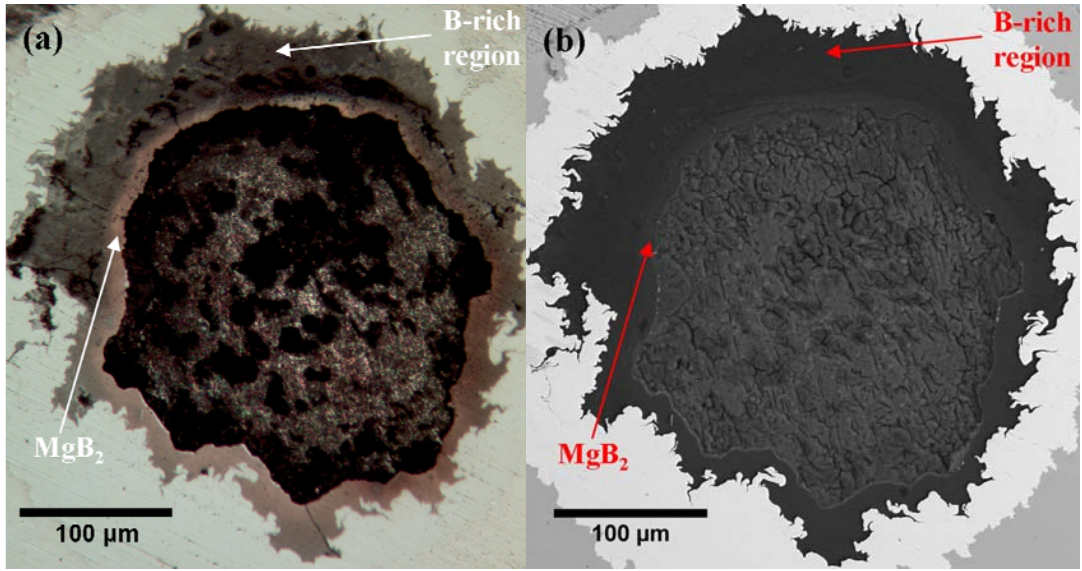

Figure. 2 (a) Optical microscopic and (b) back scattered SEM images taken from the same transverse cross section in the 3 mol% C doped AIMI strand A3.

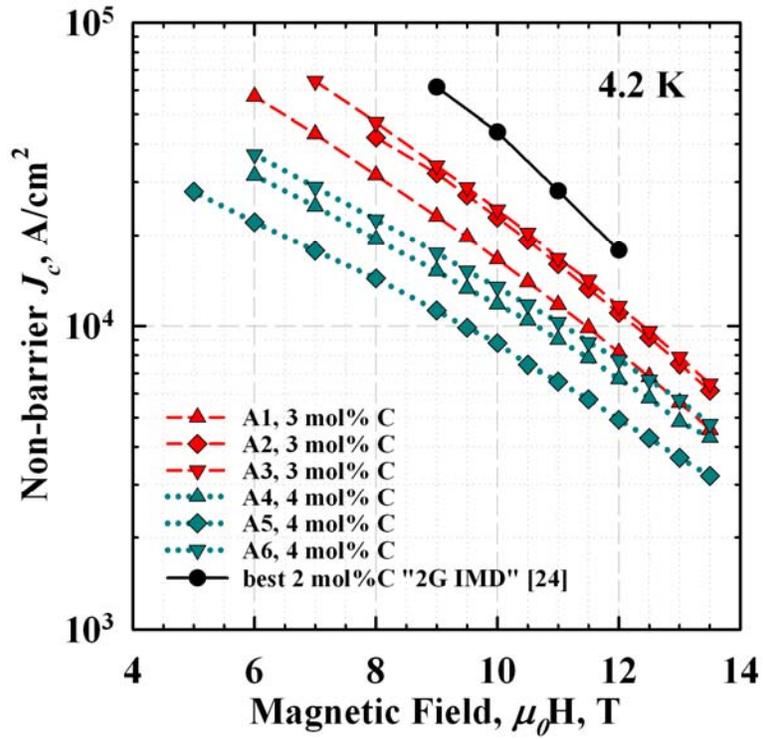

Figure. 3 Field dependence of the non-barrier $J_c$s of the 3 mol% and 4 mol% C doped monofilamentary AIMI strands at 4.2 K. A fully transformed 2 mol% C doped "2G-IMD" strand is included [24].

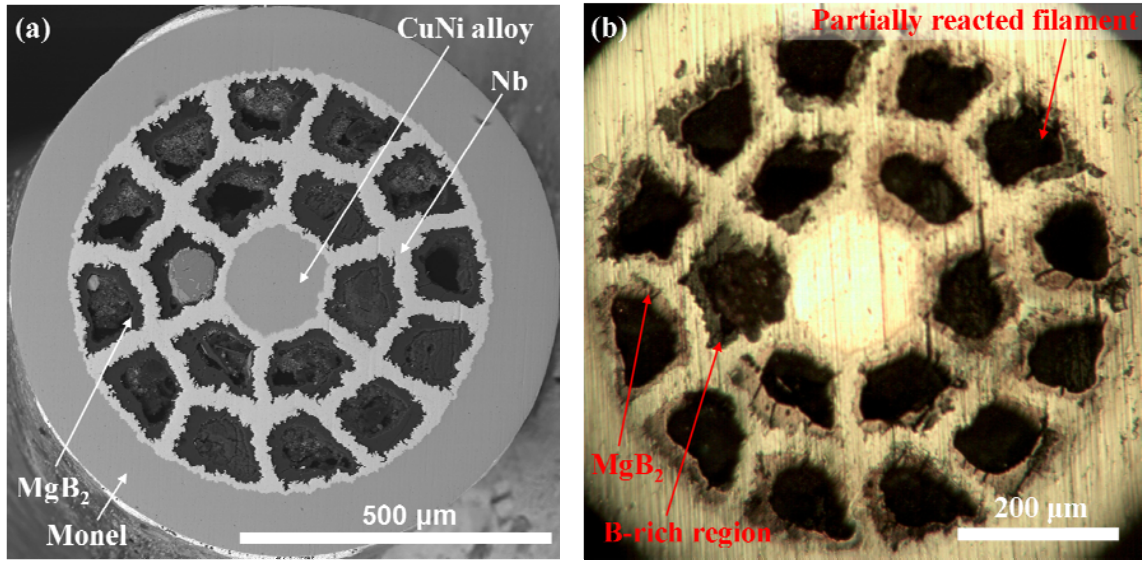

Figure. 4 Transverse cross-sectional (a) SEM and (b) OM images of the 2 mol% C doped multifilamentary AIMI strand B1.

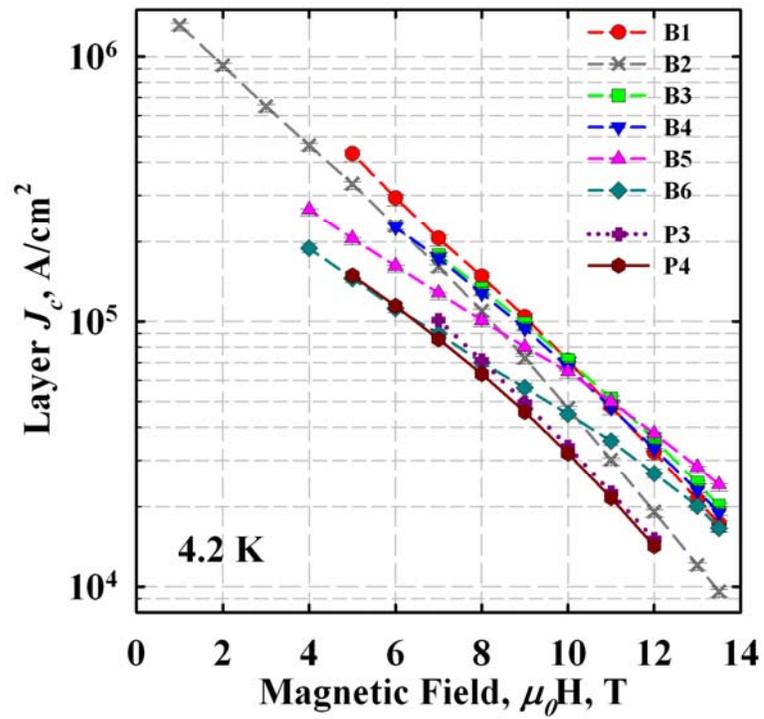

Figure 5. Field dependence of the layer $J_c$s for multifilamentary AIMI strands at 4.2 K, data from P3 and P4 are included for comparison. Here B1 and B2 are one-meter-long "ITER-barrel" type strands.

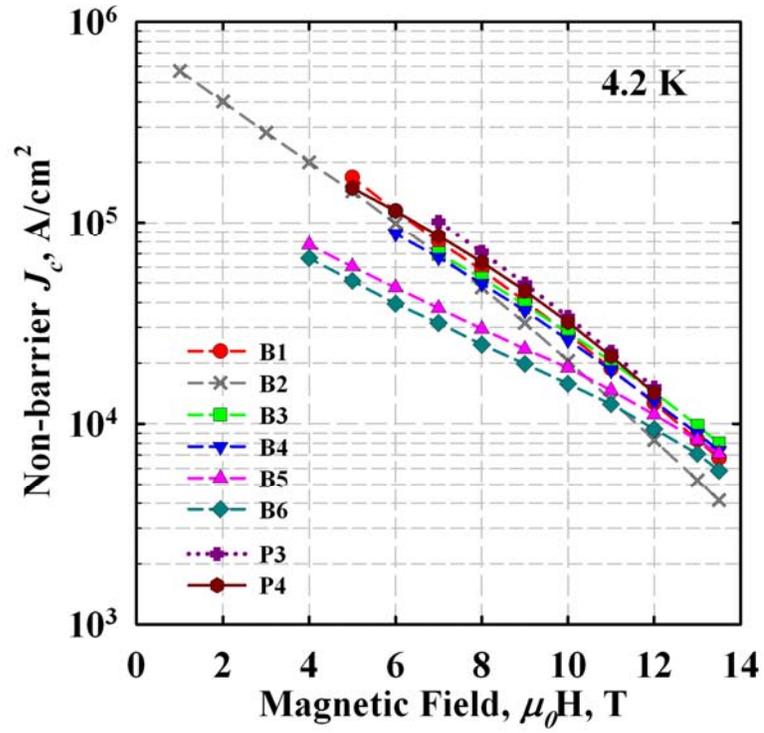

Figure 6. Field dependence of the non-barrier $J_c$s for multifilamentary AIMI strands at 4.2 K, data from P3 and P4 are included for comparison. Here B1 and B2 are one-meter-long "ITER-barrel" type strands.

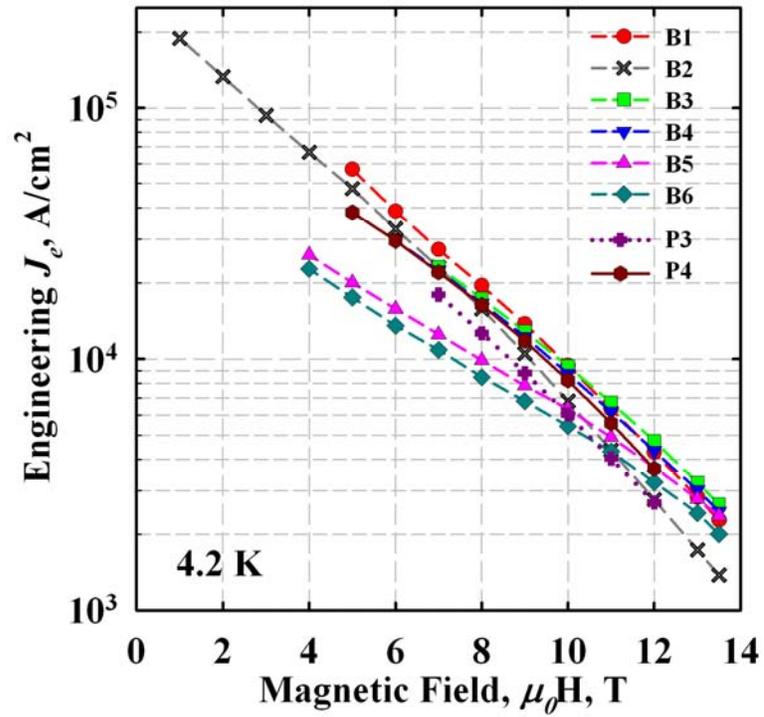

Figure 7. Field dependence of the engineering $J_e$s for multifilamentary AIMI strands at 4.2 K, data from P3 and P4 are included for comparison. Here B1 and B2 are one-meter-long "ITER-barrel" type strands.

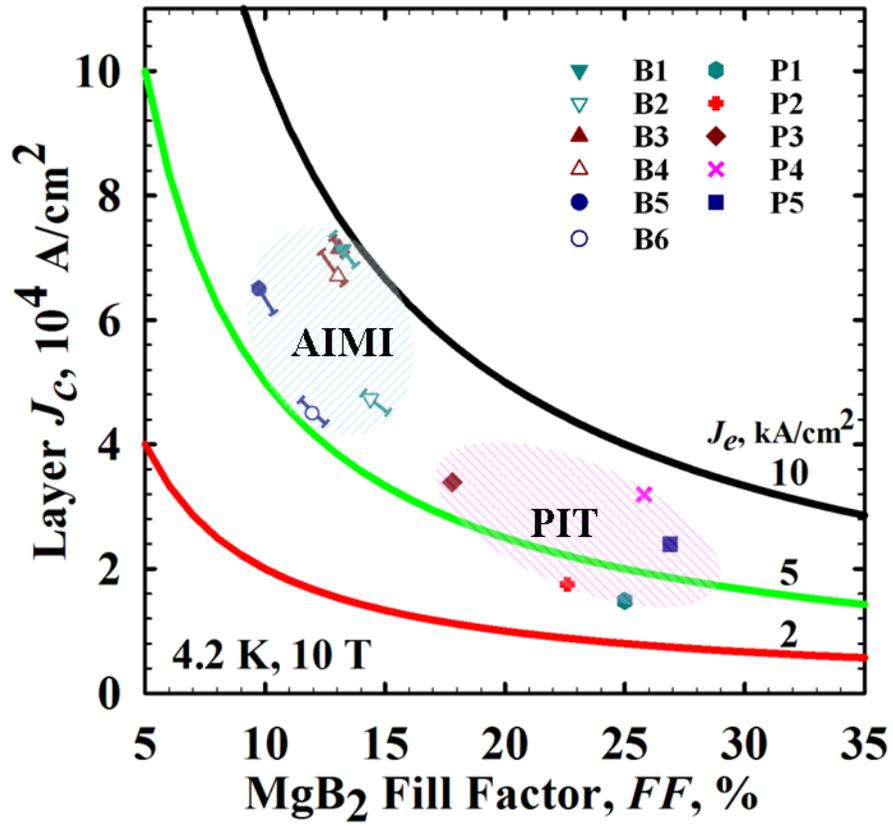

Figure 8. Comparison and relationships between $J_e$, layer $J_c$ and MgB$_2$ fill factor ($FF$) of multifilamentary AIMI strands and PIT strands with different C concentrations at 4.2 K, 10 T. Error bars are added for AIMI strands to describe the accuracy of the data.